%Paper: hep-ph/9205208
%From: sjs@ufhepa.phys.ufl.edu
%Date: Thu, 07 May 1992 21:44:53 EDT

%%%%%%%%%%%%%%%%%%%%%%%%%%%%%%%%%%%%%%%%%%%%%%%%%%%%%%%%%%%%%%%%%%%%%%%%%%%%
%
%	Use PLAIN TEX --- all macros included.
%	Postscript figures will be extracted automatically and included.
%
%	(However, the use of \special can be site-dependent so if this
%	fails to include, print out fig1.ps, fig2.ps, fig3.ps separately.)
%
%%%%%%%%%%%%%%%%%%%%%%%%%%%%%%%%%%%%%%%%%%%%%%%%%%%%%%%%%%%%%%%%%%%%%%%%%%%%

\catcode`\@=11 % This allows us to modify PLAIN macros.

%%%%%%%%%%%%%%%%%%%%%%%%%%%%%%%%%%%%%%%%%%%%%%%%%%%%%%%%%%%%%%%%%%%%%%%%%%%%
%	Macros to extract figures to separate files, a la phyzzx
%%%%%%%%%%%%%%%%%%%%%%%%%%%%%%%%%%%%%%%%%%%%%%%%%%%%%%%%%%%%%%%%%%%%%%%%%%%%

\newwrite\FIGUREwrite
\def\PSFIGURE#1{\immediate\openout\FIGUREwrite=#1 \PSFG@begin\PSFG@end}
{\catcode`\^^M=\active %
 \gdef\PSFG@begin{
\catcode`\%=11 %% This allows comments to go into output file
   \begingroup \catcode`\^^M=\active \let^^M=\relax}%
 \gdef\PSFG@end#1{\PSFG@@end #1^^M\PSFG@terminate \endgroup
    \catcode`\%=14 }% Comments are comments again
 \gdef\PSFG@@end#1^^M{\toks0={#1}\immediate\write\FIGUREwrite{\the\toks0}%
   \futurelet\N@XT\PSFG@test}%
 \gdef\PSFG@test{\ifx\N@XT\PSFG@terminate \let\N@XT=\relax%
       \else \let\N@XT=\PSFG@@end \fi \N@XT}%
}
\let\PSFG@terminate=\relax

%%%%%%%%%%%%%%%%%%%%%%%%%%%%%%%%%%%%%%%%%%%%%%%%%%%%%%%%%%%%%%%%%%%%%%%%%%%%
%	A few fonts
%%%%%%%%%%%%%%%%%%%%%%%%%%%%%%%%%%%%%%%%%%%%%%%%%%%%%%%%%%%%%%%%%%%%%%%%%%%%

%%\input fonts
\font\twelvebf=cmbx12   %cmbx10 scaled\magstep1
\font\ninerm=cmr9

%%%%%%%%%%%%%%%%%%%%%%%%%%%%%%%%%%%%%%%%%%%%%%%%%%%%%%%%%%%%%%%%%%%%%%%%%%%%
%	Numbering macros from harvmac.tex
%%%%%%%%%%%%%%%%%%%%%%%%%%%%%%%%%%%%%%%%%%%%%%%%%%%%%%%%%%%%%%%%%%%%%%%%%%%%
%
%       use \nolabels to get rid of eqn, ref, and fig labels in draft mode
\def\nolabels{\def\wrlabeL##1{}\def\eqlabeL##1{}\def\reflabeL##1{}}
\def\writelabels{\def\wrlabeL##1{\leavevmode\vadjust{\rlap{\smash%
{\line{{\escapechar=` \hfill\rlap{\sevenrm\hskip.03in\string##1}}}}}}}%
\def\eqlabeL##1{{\escapechar-1\rlap{\sevenrm\hskip.05in\string##1}}}%
\def\reflabeL##1{\noexpand\llap{\noexpand\sevenrm\string\string\string##1}}}
\nolabels
%
% tagged sec numbers
\global\newcount\secno \global\secno=0
\global\newcount\meqno \global\meqno=1
\def\newsec#1{\global\advance\secno by1%%%\message{(\the\secno. #1)}
\global\subsecno=0\eqnres@t\medskip\noindent{\it\the\secno. #1 ---}
%%%\writetoca{{\secsym} {#1}}\par\nobreak\medskip\nobreak}
\writetoca{{\secsym} {#1}}}
\def\eqnres@t{\xdef\secsym{\the\secno.}\global\meqno=1}
\def\sequentialequations{\def\eqnres@t{\bigbreak}}\xdef\secsym{}
\global\newcount\subsecno \global\subsecno=0
\def\subsec#1{\global\advance\subsecno by1\message{(\secsym\the\subsecno. #1)}
\ifnum\lastpenalty>9000\else\bigbreak\fi
\noindent{\it\secsym\the\subsecno. #1}\writetoca{\string\quad
{\secsym\the\subsecno.} {#1}}\par\nobreak\medskip\nobreak}
\def\appendix#1#2{\global\meqno=1\global\subsecno=0\xdef\secsym{\hbox{#1.}}
\bigbreak\bigskip\noindent{\bf Appendix #1. #2}\message{(#1. #2)}
\writetoca{Appendix {#1.} {#2}}\par\nobreak\medskip\nobreak}
%
%       \eqn\label{a+b=c}	gives displayed equation, numbered
%				consecutively within sections.
%     \eqnn and \eqna define labels in advance (of eqalign?)
%
\def\eqnn#1{\xdef #1{(\secsym\the\meqno)}\writedef{#1\leftbracket#1}%
\global\advance\meqno by1\wrlabeL#1}
\def\eqna#1{\xdef #1##1{\hbox{$(\secsym\the\meqno##1)$}}
\writedef{#1\numbersign1\leftbracket#1{\numbersign1}}%
\global\advance\meqno by1\wrlabeL{#1$\{\}$}}
\def\eqn#1#2{\xdef #1{(\secsym\the\meqno)}\writedef{#1\leftbracket#1}%
\global\advance\meqno by1$$#2\eqno#1\eqlabeL#1$$}
%
%
%     \ref\label{text}
% generates a number, assigns it to \label, generates an entry.
% To list the refs on a separate page,  \listrefs
%
\global\newcount\refno \global\refno=1
\newwrite\rfile
\def\ref{[\the\refno]\nref}
\def\nref#1{\xdef#1{[\the\refno]}\writedef{#1\leftbracket#1}%
\ifnum\refno=1\immediate\openout\rfile=refs.tmp\fi
\global\advance\refno by1\chardef\wfile=\rfile\immediate
%%%\write\rfile{\noexpand\item{#1\ }\reflabeL{#1\hskip.31in}\pctsign}\findarg}
\write\rfile{\noexpand\item{$^{#1}$}\reflabeL{#1\hskip.31in}\pctsign}\findarg}
%	horrible hack to sidestep tex \write limitation
\def\findarg#1#{\begingroup\obeylines\newlinechar=`\^^M\pass@rg}
{\obeylines\gdef\pass@rg#1{\writ@line\relax #1^^M\hbox{}^^M}%
\gdef\writ@line#1^^M{\expandafter\toks0\expandafter{\striprel@x #1}%
\edef\next{\the\toks0}\ifx\next\em@rk\let\next=\endgroup\else\ifx\next\empty%
\else\immediate\write\wfile{\the\toks0}\fi\let\next=\writ@line\fi\next\relax}}
\def\striprel@x#1{} \def\em@rk{\hbox{}}
\def\lref{\begingroup\obeylines\lr@f}
\def\lr@f#1#2{\gdef#1{\ref#1{#2}}\endgroup\unskip}

\def\addref#1{\immediate\write\rfile{\noexpand\item{}#1}} %now unnecessary
\def\startrefs#1{\immediate\openout\rfile=refs.tmp\refno=#1}
\def\xref{\expandafter\xr@f}\def\xr@f[#1]{#1}
\def\refs#1{\count255=1[\r@fs #1{\hbox{}}]}
\def\r@fs#1{\ifx\und@fined#1\message{reflabel \string#1 is undefined.}%
\nref#1{need to supply reference \string#1.}\fi%
\vphantom{\hphantom{#1}}\edef\next{#1}\ifx\next\em@rk\def\next{}%
\else\ifx\next#1\ifodd\count255\relax\xref#1\count255=0\fi%
\else#1\count255=1\fi\let\next=\r@fs\fi\next}
%

%
% this is ugly, but moore insists
\newwrite\ffile\global\newcount\figno \global\figno=1
\def\fig{fig.~\the\figno\nfig}
\def\nfig#1{\xdef#1{fig.~\the\figno}%
\writedef{#1\leftbracket fig.\noexpand~\the\figno}%
\ifnum\figno=1\immediate\openout\ffile=figs.tmp\fi\chardef\wfile=\ffile%
\immediate\write\ffile{\noexpand\medskip\noexpand\item{Fig.\ \the\figno. }
\reflabeL{#1\hskip.55in}\pctsign}\global\advance\figno by1\findarg}
\def\xfig{\expandafter\xf@g}\def\xf@g fig.\penalty\@M\ {}
\def\figs#1{figs.~\f@gs #1{\hbox{}}}
\def\f@gs#1{\edef\next{#1}\ifx\next\em@rk\def\next{}\else
\ifx\next#1\xfig #1\else#1\fi\let\next=\f@gs\fi\next}
\newwrite\lfile
{\escapechar-1\xdef\pctsign{\string\%}\xdef\leftbracket{\string\{}
\xdef\rightbracket{\string\}}\xdef\numbersign{\string\#}}

\def\writestop{\def\writestoppt{\immediate\write\lfile{\string\pageno%
\the\pageno\string\startrefs\leftbracket\the\refno\rightbracket%
\string\def\string\secsym\leftbracket\secsym\rightbracket%
\string\secno\the\secno\string\meqno\the\meqno}\immediate\closeout\lfile}}
\def\writestoppt{}\def\writedef#1{}
\def\seclab#1{\xdef #1{\the\secno}\writedef{#1\leftbracket#1}\wrlabeL{#1=#1}}
\def\subseclab#1{\xdef #1{\secsym\the\subsecno}%
\writedef{#1\leftbracket#1}\wrlabeL{#1=#1}}
\newwrite\tfile \def\writetoca#1{}
\def\leaderfill{\leaders\hbox to 1em{\hss.\hss}\hfill}
%	use this to write file with table of contents
\def\writetoc{\immediate\openout\tfile=toc.tmp
   \def\writetoca##1{{\edef\next{\write\tfile{\noindent ##1
   \string\leaderfill {\noexpand\number\pageno} \par}}\next}}}
%       and this lists table of contents on second pass

%

%%%%%%%%%%%%%%%%%%%%%%%%%%%%%%%%%%%%%%%%%%%%%%%%%%%%%%%%%%%%%%%%%%%%%%%%%%%%
%	Double and single columns a la K.McD.
%%%%%%%%%%%%%%%%%%%%%%%%%%%%%%%%%%%%%%%%%%%%%%%%%%%%%%%%%%%%%%%%%%%%%%%%%%%%

\ \tolerance=10000
\parindent=10pt
\parskip=0pt
\def\today{\ifcase\month\or
   January\or February\or March\or April\or May\or June\or
   July\or August\or September\or October\or November\or December\fi
   \space\number\day, \number\year}
%%%%%%%%%%%% setup for 54 char./line, and 3 lines/inch
%\magnification=\magstep1
%\hsize=4.05truein
%\hoffset=1.1truein
%\voffset=0truein
%\baselineskip=20pt
%%%%%%%%%%%%%%% set-up for P.R.L. facsimile
%% similiar to macros in Texbook, p 415 ff.
\newdimen\pagewidth \newdimen\pageheight \newdimen\textheight
\newdimen\headheight \newdimen\footheight
\hsize=7.0truein
\hoffset=-0.25truein
\vsize=10.2truein % 732pt
\voffset=-0.6truein
\baselineskip=12pt
\pagewidth=\hsize \pageheight=\vsize
\headheight=32pt
\textheight=672pt % 56 lines of 12pt each
\footheight=20pt
\maxdepth=2pt
%% define the PRL header
  %cmbx10 scaled\magstep1

\font\largeheadfont=cmcsc10
\headline={\hbox to\pagewidth{%
%{\smallheadfont Volume 1066, Number 29}\hskip45pt%
%{\largeheadfont PHYSICAL REVIEW D}\hfil%
%{\smallheadfont 1 April 2001}}}
{\largeheadfont University of Florida}
\hfil{\largeheadfont UFIFT-HEP-92/11}}}
%% define the PRL footer
\nopagenumbers % page numbers handled in the footer
\footline={\hbox to\pagewidth{\hskip125pt%
%\copyright\enskip 2001 The American Physical Society%
% Submitted to Physical Review Letters%
\hfil\folio}}
\def\evenpage{\hbox to\pagewidth{\folio\hfil}}
\def\oddpage{\hbox to\pagewidth{\hfil\folio}}
%%%%%
\def\onepageout#1{\shipout\vbox{
      \offinterlineskip % group 3 boxes into 1 page
      \vbox to\headheight{\the\headline \vskip10pt \hrule height1pt \vfill}
      \vbox to\textheight{#1} % text goes in here
      \vbox to\footheight{\vfil
              \ifnum\pageno=1\the\footline \else{\ifodd\pageno\oddpage
              \else\evenpage \fi}\fi}}
      \advancepageno}
\newbox\partialpage
\def\begindoublecolumns{\begingroup
       \output={\global\setbox\partialpage=\vbox{\unvbox255\bigskip}}\eject
       \output={\doublecolumnout} \hsize=3.375truein \vsize=1356pt}
\def\enddoublecolumns{\output={\balancecolumns}\eject
       \endgroup \pagegoal=\vsize}
\def\doublecolumnout{\splittopskip=\topskip \splitmaxdepth=\maxdepth
       \dimen@=672pt \advance\dimen@ by-\ht\partialpage
       \setbox0=\vsplit255 to\dimen@ \setbox2=\vsplit255 to\dimen@
       \onepageout\pagesofar \unvbox255 \penalty\outputpenalty}
\def\pagesofar{\unvbox\partialpage
       \wd0=\hsize \wd2=\hsize \hbox to\pagewidth{\box0\hfil\box2}}
\def\balancecolumns{\setbox0=\vbox{\unvbox255} \dimen@=\ht0
       \advance\dimen@ by\topskip \advance\dimen@ by-\baselineskip
       \divide\dimen@ by2 \splittopskip=\topskip
       {\vbadness=10000 \loop \global\setbox3=\copy0
           \global\setbox1=\vsplit3 to\dimen@
           \ifdim\ht3>\dimen@ \global\advance\dimen@ by1pt \repeat}
       \setbox0=\vbox to\dimen@{\unvbox1} \setbox2=\vbox to\dimen@{\unvbox3}
       \pagesofar}

\catcode`\@=12 % at signs are no longer letters

%%%%%%%%%%%%%%%%%%%%%%%%%%%%%%%%%%%%%%%%%%%%%%%%%%%%%%%%%%%%%%%%%%%%%%%%%%%%
%	 Figures begin here
%        %%%%%%%%%%%%%%%%%%
%          end of figure
%%%%%%%%%%%%%%%%%%%%%%%%%%%%%%%%%%%%%%%%%%%%%%%%%%%%%%%%%%%%%%%%%%%%%%%%%%%%
%% other changes to harvmac
\def\Title#1{\centerline{\twelvebf #1}\bigskip}

\def\authornote#1#2{{#1}}
\def\footnote#1#2{\ref\NONE{#2}} %% Mix footnotes with references

\def\ABSTRACT#1{
{\ninerm
\centerline{(\DATE)}
\smallskip \parindent=.75in
{\narrower \parindent=10pt
{#1}
\medskip}}
\parindent=10pt
\begindoublecolumns}

%%\input hmbody

%%%%%%%%%%%%%%%%%%%%%%%%%%%%%%%%%%%%%%%%%%%%%%%%%%%%%%%%%%%%%%%%%%%%%%%%%%%%
%       Text begins here
%%%%%%%%%%%%%%%%%%%%%%%%%%%%%%%%%%%%%%%%%%%%%%%%%%%%%%%%%%%%%%%%%%%%%%%%%%%%

% Misc. definitions
\def\OMIT#1{}

\def\ccdot{\hbox{\kern-.1em$\cdot$\kern-.1em}}

\def\art#1{{\sl #1}}
\def\art#1{}

\def\DATE{April 1992}
\Title{{%
Late time Cosmological Phase transition and Galactic Halo as Bose-Liquid
}}
\centerline{Sang-Jin Sin\authornote{$^\dagger$}%
{Department of Energy under contract DE-FG05-86ER-40272.
\hfill\break
email: sjs@ufhepa.phys.ufl.edu}
}
\smallskip
\centerline{\it Department of Physics, University of Florida,
   Gainesville FL.32611}
%\smallskip
\ABSTRACT{
We consider the ultra light pseudo Nambu-Goldstone boson
appearing  in the late time cosmological phase transition theories
as a dark matter candidate. Since it is almost massless, its
nature is  more wave like than particle like. Hence
we apply quantum mechanics to study how they form the galactic halos.
Three predictions are made; (1)the mass profile $\rho\sim r^{-1.6}$,
(2)there are ripple-like fine structures in rotation curve,
(3) the rotation velocity times ripple's wave length is
largely galaxy independent.
We compare the rotation curves predicted by our theory with the data observed.
}

\def\d{\partial}
\def\r{{ \hat r}}
\def\del{\Delta}
\def\M{{\hat M}}
\def\hpsi{{\hat \psi}}
\def\ep{\epsilon}

\newsec{Introduction}
Recently, motivated by the large scale structure, an ultra light
Nambu-Goldstone boson  was
introduced in late time cosmological phase transition
theory.\ref\hill{C.T.Hill, D.N. Schramm, J. Fry,
      Comments of Nucl.\& Part. Phys.{\bf 19},25.}
The bottom line of the theory is that if a phase transition happens {\it after}
decoupling, one can avoid the constraint imposed by
isotropy of microwave back ground.
If an event happens so late then the universe is already big, hence
the relevant particle whose compton wave length provides
 the scale of interesting
structure must be very light and
this particle could be a dark matter candidate.
\ref\frieman{J. Frieman, C. Hill and R. Watkins, Fermilab-PUB-91/324-A. }

The dark matter distribution is most clearly imprinted in the rotation curve.
The overall $`$flatness' of the rotation curve (RC) is equivalent to
mass profile $\rho \sim r^{-2}$, which can be obtained
by assuming a thermal distribution.
\ref\lindenbell{ D. Linden-Bell, MNRAS,(1967) {\bf 136},101.}
\ref\king{I. R. King, Astron. J.   {\bf71}64(1966).}
Another way suggested to understand the dark matter distribution is
a spherical infall model
\ref\gott{J. Gott, Ap. J. {\bf 201},296.}
\ref\gunn{J. Gunn, Ap. J.   {\bf 218 }592.}
\ref\filmore{J. Filmore and P. Goldreich {\bf 281 }1.}
\ref\bertschinger{E. Bertschinger, Ap. J. Suppl.   {\bf58} 39. }
\ref\sikivie{ P. Sikivie and J. Ipser, UFIFT-HEP-91-17}
, where the mass profile is given
by $\rho\sim r^{-2.25}$, giving a slowly decreasing rotation curve.
However, the observed rotation curves are
actually slightly increasing
\ref\rubin{V. Rubin et.al, Ap. J. {\bf 289 } 81 (1985);
Ap. J. {\bf 261 } 439(1982); Ap. J. {\bf 238 } 471 (1980).}
though they are said to be
$`$flat'.
So neither the thermal distribution nor the infall model is
completely satisfactory.

If dark matter  consists of these particles whose mass is, say, $m\sim
10^{-24}eV$ and dark matter density around us is
$\sim 10^{-25}g/cm^3$,
then the inter-particle distance is order of $10^{-13}cm$, while the
compton wave length is of 10pc order. Hence there is no point talking
about the individual particle's position and momentum. The
dark matter is more like a wave than a particle,
and the galactic halos are giant systems of condensed bose liquid.
Here we are in the region reigned by quantum mechanics.
In this context, we will consider the dark matter distribution
quantum mechanically and this is the purpose of our paper.
Our result will show mass profile $\rho\sim r^{-1.6}$ leading to
slightly increasing rotation curve.
Further more due to the system's wave nature,
rotation curve has the ripple-like fine structure.
By the uncertainty principle, our theory predicts that
the rotation velocity times the  ripple's wave length is
galaxy independent.

\newsec{Quantum considerations of dark matter}
We consider the spherically symmetric case for  simplicity and also
assume that there are no inter-particle interactions apart from gravity.
The classical energy of a particle moving inside a potential generated
by other particles is
$$E={1\over 2} mv^2 + \int_0^r dr'{GmM(r')\over r'^2 },\,\,
\eqno(1)$$
$ {\rm where }\,
M(r)=\int_0^r dr' 4\pi {r'}^2 \rho(r') $.
Quantum mechanically,
we are interested in a wave equation which governs the collective behavior
of highly correlated bosonic particles.
In principle, one has to work out Hartree's self consistent procedure,
but here the fact that the system is
a highly correlated system of bosons simplifies the life.
Assuming that all or significant fraction
of particles are in the same state,  we take the condensation wave
function $\psi$
\ref\landau{Landau and Lifshitz, Course of theoretical physics
Vol.9 section 26.}
as the Hartree's self-consistent field.
Then the density of particles is proportional to $|\psi|^2$ and we obtain
a wave equation given by
$$ i\hbar\d_t \psi=-{\hbar^2\over 2m}\del \psi+ GmM_0
\!\int_0^r\! {dr' \over r'^2 }\!\int_0^r dr'' 4\pi{r''}^2 |\psi|^2 \cdot
\psi(r) \eqno(2)$$
Here $M_0$ in $M=M_0\int dr 4\pi{r}^2 |\psi|^2$
is a mass parameter introduced for dimensional reasons.
The equation is non-linear, hence we can not superpose two solutions, nor do we
have freedom to normalize it.
However, as we shall show later, this equation has a scaling symmetry
which allows us to resolve the ambiguity due to the choice of $M_0$. That is,
the physical quantity $M$ does not depend on the choice of $M_0$, which is a
nontrivial property.

Another remark is  on the non-relativistic treatment.
(From now on we set $\hbar=c=1$ except for emphasis.)
One might wonder that when the mass of the particle is so small,
whether a non-relativistic treatment is correct. The justification is two-fold.
The basic reasons are (1)the virial theorem tells us $v^2 = GM/r$ is
independent of
individual particle mass m and (2)the total mass of the system is small enough
as we shall show below.
 When the effective coupling constant $e^2:= GmM\sim 1$ or equivalently
when  $M\sim 1/Gm$, the gravitational Bohr radius $r_0\sim
1/me^2$ is equal to the compton wave length $1/m$, and the size of the system
is equal to the size of a particle. This means that
particles lie on top of one another, which is nothing but a black hole
system.
In fact,  the Schwarzschild radius is $2GM \sim 1/m$, confirming
our picture.  When this happens, the non-relativistic treatment breaks down.
In our case, however,
$1/Gm\sim 10^{14}M_\odot$ while $M\sim 10^{12}M_\odot$ so that
$v \sim e^2 \sim 10^{-2}$. Hence the non-relativistic treatment is well
justified.

We are interested in a {\it stationary }
solution $\psi(r,t)= e^{-iEt/\hbar}\psi(r)$.
After scaling
$$r= r_0 \r,\,\,\, \psi =r_0^{-3/2}\hpsi,\,\,\,
E={\hbar^2\ep\over 2m}, \, \,\, \eqno(3)$$
 where $r_0=\hbar^2/2GM_0m^2$,
we can rewrite (2) in terms of the radial wave function $u(r)$,
$$ u''(\r)+\big(
\ep- \int_0^\r d\r'{ 1\over \r'^2 }\int_0^{\r'} d\r'' u^2(\r'') \big)
u(\r)=0, \eqno(4)$$
where $\hpsi(\r)={1\over\sqrt{4\pi}}{u(\r)\over \r}$.
We have solved this equation numerically.
Figures 1(a)-(b) are the stationary solutions and corresponding velocity curves
${\hat v}(r)=v(r)/v_0 = (\int_0^\r u^2 d\r'/\r) ^{1/2}$ where
$v_0=\sqrt{GM_0/r_0}$.
\vskip.1in
\hrule
\vskip.1in
\centerline{Figure 1 (a),(b)}
\vskip.1in
\hrule
\vskip.1in
We are interested in a {\it higher state} with nodes
rather than the ground state,
because there is no way for the system to lose its energy
to relax to its ground state, since there is no dissipation mechanism.
Let's define the node number $n$ as the number of nodes plus 1.
For $n$ greater than 4, the rotation curve
resembles the observed ones and the characteristic features
do not change as we increase the number of nodes.
The numerical study show that mass profile $\rho\sim r^{-1.56}, r>>1$ for n=6,
and the power depends on n very weakly.

We can $`$understand' (not derive)
 behaviors qualitatively by simple semi-classical analysis
\ref\wkb{ A cautious remark on terminology;
There is no WKB method for non-linear equations in general.
Here the non-linearity is contained only in potential that is
obtained numerically. Hence we are applying WKB to effectively
linear equations.
Since the solution's behaviour is such that
the resulting potential varies slowly, WKB is a good approximation.
}
Since $M(r)\sim rv^2$, flatness of rotation curve
is equivalent to $M'(r)\sim u^2$ being constant.
Usual WKB method gives
$$ u(r)\sim {1\over \sqrt{p_r}}\cos (\int_0^r p_r dr - C\pi),
\eqno(5) $$
where $p_r=\sqrt{2m(\ep-V(u))}$.
Since the potential is increasing  very slowly
as a function of $r$, so is $1/\sqrt{p_r}$; hence it gives a minor modification
to the $`$over-all flat' cosine curve.
This explains the overall flat but slowly increasing aspect of the rotation
curves.
Also notice that the last wave is almost as long as the sum of all the others
since $p_r$ near the $`$turning point' is singular.
The ripple like structure is an inevitable phenomena predicted in this
approach.
It is a direct consequence of using the wave theory
under the one state (or small number of states) dominance asumption.

Apart from mass profile and ripple's existence, our theory has one
quantitative prediction which could prove or disprove it
with more precise measurements of rotation curves.

\noindent
{\it Rotation velocity times the ripple wave length is largely
 Galaxy independent.}

\noindent{This is nothing but re-phrasing the uncertainty principle.
More precisely,}
$$ v_{rotation}\lambda_{ripple}=
{1\over \sqrt{2}}{1\over m} f(\r) ,\eqno(6)$$
where $f(\r)$ is a slowly increasing dimension-less function
given by
\def\V{{\hat V}}
$$f(\r)=\sqrt{\int_0^\r d\r u^2(\r)\over \r (\V(\r_{t})-\V(\r))} ,\eqno(7)$$
where $\V(\r)=V(r)/mv_0^2$ and $r_{t}$ is
$r_{t}$ is  the effective classical turning point.
If we further assume that the potential changes very slowly
as we change $n>>1$, we get the eigenvalue distribution
%$\ep_n \sim \ln (n+3/4)$ for log potential, and
$\ep_n\sim \ep_0+ (n+\beta)^{2\alpha /(2+\alpha)}$ for $V\sim r^\alpha$.
Numerical study shows that without including visible matter, $\alpha=0.44$ for
n=6, and 0.45 for n=5.
When we compare with data we have to include visible matter and this leads
to $\alpha\sim 0.39$. See figure 3(c).

\vskip.1in
\hrule
\vskip.1in
\centerline{Figure 2 (a),(b)}
\vskip.1in
\hrule
\vskip.1in

Next we observe that as we vary the initial condition $\hpsi_0$,
the scale free quantities $\ep$ and $\M:=\int u^2 dr$  have a simple
dependence on it.  Namely
$$\M= 20.3068 \sqrt{\hpsi_0}, \,\, \ep= 1.4900\hpsi_0 \,\,\,{\rm for}\,\,n=5
\eqno(8)$$
The coefficients depend on n, e.g, for n= 6 (5 nodes), they are given
by 24.4408 and 1.5385 respectively.
To understand this we have proved that equation (2) has the following symmetry.

{\bf Scaling symmetry} : {\it if $\psi(r,\ep)$ is a solution, so is $\lambda^2
\psi(\lambda r,\lambda^2\ep)$  }

\noindent The relation (5) is just a corollary of this rather interesting
property.
Also, by choosing $\lambda$ such that the norm of $\psi$ is equal to one,
one can equate $M_0$ and $M$.  This is not convenient in practical
numerical integration, however. We will discuss this property when we
discuss the data fitting.

\newsec{ Comparison with observations}
We now compare the the rotation curves predicted in our theory
and the actual observations.
First we have to include the visible matter's effect. See figure 3(a),(b).
Equation (2) has to be modified to be
$$ i\d_t \psi= {-1\over 2m}\del \psi+
Gm\int_0^r\! { dr'\over r'^2 }\int_0^r dr'' 4\pi(M_0 |\psi|^2 +\rho_{vis})
\cdot \psi\eqno(9)$$
We choose the effective density of visible matter as the
Plumber's potential for the bulge and a Yukawa type exponential decay for the
disk.
By including the visible matter's contribution, the rotation curve becomes more
flat, but the wavelength of the ripple does not change.
\vskip.1in
\hrule
\vskip.1in
\centerline{Figure 3 (a),(b),(c),(d)}
\vskip.1in
\hrule
\vskip.1in
The ripple structure is  manifest for relatively few galaxies.
The data of Rubin et.al
 showed that in any galaxy there are fluctuations that could indicate
the fine structure, but most of the data
is plagued with error bars that wash out the signal of the fine structure.
The data is good enough for the rotation velocity's magnitude but too crude
for the fine structure. Among 63 galaxies listed in
\refs{\rubin},
there were about five galaxies which have relatively small error bars
and show ripples. These are NGC2998, NGC1357, NGC4594, NGC1620, NGC801.
Galaxy NGC2998's data is particularly interesting for us and we plot the
data and the theoretical curve in Figure 3(d).
There are four ripples in the NGC2998's data but we suspect that out side the
measured region there can still be density peaks and nodes
that could lead to further ripples.
However as n increases, the total galaxy mass increases;
and for large enough n,
$\Omega$ exceeds 1, violating bias of present
theoretical community. For example, if n is larger than 7, the
total mass of the halo is bigger
than $10^{13}M_\odot$.  Hence there is some restriction on n.
Here we take  the minimal choice n=5 (four nodes).

By measuring the ripple wavelength and the rotation velocity,
we can deduce the
total mass M of the Galaxy and the constituent particle mass m.
For convenience we  choose $\hpsi_0=0.2$.
Comparing the velocity and wavelength of, say,  the second ripple,
we have two relations,
$$7r_0=8kpc,\,\,{\rm and} \; 0.32 v_0=204 km/sec. \eqno(10)$$
Solving these two, we get
$$m=3.3\times 10^{-23}eV ,\,\, {\rm and }\,\,M_0=0.69\times 10^{12}M_\odot
.\eqno(11)$$
 Using the numerically calculated value $\M=8.73575$,
 $$M=\M \cdot M_0=5.9\times 10^{12}M_\odot\eqno(12)$$
The central density of dark matter is given by
$$\rho_c={1/4\pi}|\hpsi_0|^2(M_0/r_0^3)=1.08\times 10^{-22}g/cm^3 .\eqno(13)$$
Therefore we can weigh the Galaxy  by looking at the rotation curve only!
In the model adopted,
the visible matter's total mass is $2.7\times 10^{11}M_\odot$.
For other galaxies, by similar analysis
the prediction of eq. (8) is roughly satisfied.
More massive galaxies have larger rotational velocity
hence  shorter ripple wavelength.
For future data analysis we give the general formula for $m,M$ with n=5.
$$m={8kpc\over\lambda}{204km/sec\over v}\times 2.7\times 10^{-23}eV \eqno(14)$$
$$M= {\lambda\over8kpc}({v\over 204km/sec})^2\times 5.9\times 10^{12}M_\odot.
\eqno(15)$$

Now we show that above  quantities are independent of the choice
of the  initial condition $\hpsi_0$ or equivalently
independent  of the choice of $M_0$.
Recall that under the scaling,
$\hpsi\to  \lambda^2\hpsi$, $\r\to \r/\lambda$ and
$\ep\to \lambda^2\ep$ must follow. Therefore
had we chosen $\lambda^2 \hpsi_0$ instead of $\hpsi_0$ above,
we would get $7kpc/{\lambda}=8km/sec$ and
$0.32v_0 {\lambda}=200km/sec$, so that we get the same $m$ and $1/\lambda$
times $M_0$ above. On the other hand $\M=\int d^3\r |\hpsi|^2 $ must scale
by $\lambda^4/\lambda^{3}$, therefore
$M=\M\cdot M_0$ is invariant under the scaling.
The same argument shows $\rho_c$ is scale free. Hence we can choose large
$\hpsi_0$ to save computing time.
Lastly, had we chosen n larger than 7, total mass for halo is larger
than $10^{13}M_\odot$.

\newsec{Conclusion}
In this paper we investigated the consequence
of abundant existence of Nambu-goldstone boson appearing in late time phase
transition.
Since the particle wavelength is far larger than the inter-particle distances,
galactic halos made of this species are highly
correlated bose liquid. We set up a version of non-linear shr\"odinger equation
describing the collective behavior of the system.
Three predictions are made; (1)mass profile $\rho\sim r^{-1.6}$,
(2)there are ripple-like fine structures in rotation curve,
(3) rotation velocity times ripple's wave length is largely galaxy independent.
We compare the rotation curves predicted by our theory with
the data observed.
We fine-tuned the mass of the particle to fit the data for halo and determine
the total mass of the galaxy NGC2998.
Among 62 rotation curve data of Rubin et.al.,
there is a one curve where all the details of the
predictions are realized,  but most of other
data is not accurate enough to definitely confirm the the theory.
We suggests that more precise measurement of rotation curve is highly
worth-while to be done.
If future observation shows
the ripple structures in many galaxies in the way predicted  in this paper, it
would be a strong indication that late-time phase transition is real.

\noindent{\it Acknowledgements} ---
I would like to thank J. Fry, P. Griffin, J. Hong, Z. Qiu, P. Ramond
for discussions,  P. Kumar and C. Thorn for conversations on bose consensation,
J. Ipser for encouragement, especially P. Sikivie for getting me interested in
this area as well as many suggestions.

%
%\eqn\fdefined{
%   \vev{0|\bar q \gamma^\mu\gamma^5 Q |B(p)}= f_B p^\mu~.
%   }
%%%%%%%%%%%%%%%%%%%%%%%%%%%%%%%%%%%%%%%%%%%%%%%%%%%%%%%%%%%%%%%%%%%%%%%%%%%%
%	References
%%%%%%%%%%%%%%%%%%%%%%%%%%%%%%%%%%%%%%%%%%%%%%%%%%%%%%%%%%%%%%%%%%%%%%%%%%%%

\smallskip
\hbox to \hsize{\hss\vrule width2.5cm height1pt \hss}
\smallskip

{\ninerm
$^\dagger$Department of Energy under contract DE-FG05-86ER-40272.
\hfill\break
email: sjs@ufhepa.phys.ufl.edu
\hfill\break
 \immediate\closeout\rfile \input refs.tmp 
}
\enddoublecolumns
\output{\onepageout{\unvbox255}}
\vfill\eject\end